\documentclass{aastex}
\usepackage{emulateapj5}
\usepackage{apjfonts}

\newcommand{\Ob} {\ensuremath{\Omega_b}}
\newcommand{\Oc} {\ensuremath{\Omega_{cdm}}}
\newcommand{\Ol} {\ensuremath{\Omega_\Lambda}}
\newcommand{\Om} {\ensuremath{\Omega_m}}
\newcommand{\Ot} {\ensuremath{\Omega_{tot}}}
\newcommand{\Od} {\ensuremath{\Omega_{\Delta}}}
\newcommand{\Obh}{\ensuremath{\Omega_bh^2}}
\newcommand{\Och}{\ensuremath{\Omega_{cdm}h^2}}
\newcommand{\ob} {\ensuremath{\omega_b}}
\newcommand{\oc} {\ensuremath{\omega_{cdm}}}
\newcommand{\h}  {\ensuremath{h}}
\newcommand{\tc} {\ensuremath{\tau_c}}
\newcommand{\ns} {\ensuremath{n_s}}
\newcommand{\Cten}{\ensuremath{\mathcal{C}_{10}}}

\newcommand{\cmbfast}{CMBFAST}
\newcommand{\dasi}   {DASI}
\newcommand{\dmr}    {DMR}

\newcommand{\tcp}  {\ensuremath{0.0\leq \tau_c \leq0.4}}

\newcommand{\Otc}  {\ensuremath{\Omega_{tot}=1.04\pm0.06}}
\newcommand{\nsc}  {\ensuremath{n_s=1.01^{+0.08}_{-0.06}}}

\newcommand{\Obhc} {\ensuremath{\Omega_bh^2=0.022^{+0.004}_{-0.003}}}
\newcommand{\Ochc} {\ensuremath{\Omega_{cdm}h^2=0.14\pm0.04}}

\newcommand{\Otcsh} {\ensuremath{\Omega_{tot}=1.00\pm0.04}}
\newcommand{\Omc}  {\ensuremath{\Omega_m=0.40\pm0.15}}
\newcommand{\Olc}  {\ensuremath{\Omega_\Lambda=0.60\pm0.15}}

\newcommand{\nsct}  {\ensuremath{n_s=0.97^{+0.05}_{-0.04}}}

\newcommand{\bfm}{$(\Om,\Ol,\Obh,\Och,\tc,\ns,\Cten)=
(0.725,0.325,0.0200,0.15,0.0,0.95,800)$}
\newcommand{\bfma}{$(\Ob,\Oc,\Ol,\tc,\ns,\h)=(0.09,0.64,0.33,0,0.95,0.48)$}

\begin{document}

\slugcomment{Published in the Astrophysical Journal, 568, 46}

\title{Cosmological Parameter Extraction from the
First Season of Observations with DASI}

\author{C.\ Pryke, N.\ W.\ Halverson, E.\ M.\ Leitch,
J.\ Kovac, J.\ E.\ Carlstrom}
\affil{University of Chicago,
5640 South Ellis Ave.,
Chicago, IL 60637}
\author{W.\ L.\ Holzapfel}
\affil{University of California,
426 Le Conte Hall,
Berkeley, CA 94720}
\author{M.\ Dragovan}
\affil{Jet Propulsion Laboratory,
California Institute of Technology,
4800 Oak Grove Drive,
Pasadena, CA 91109}

\begin{abstract}
The Degree Angular Scale Interferometer (\dasi) has measured the
power spectrum of the Cosmic Microwave Background anisotropy over the
range of spherical harmonic multipoles  $100<l<900$.
We compare this data, in combination with the COBE-DMR
results, to a seven dimensional grid of adiabatic CDM models.
Adopting the priors $h>0.45$ and \tcp,
we find that the total density of the Universe \Otc,
and the spectral index of the initial scalar fluctuations \nsc,
in accordance with the predictions of inflationary theory.
In addition we find that the physical density of
baryons \Obhc, and the physical density of cold dark matter \Ochc.
This value of \Obh\ is consistent with that derived
from measurements of the primeval deuterium abundance
combined with big bang nucleosynthesis theory.
Using the result of the HST Key Project, $\h=0.72\pm0.08$, we find
that \Otcsh, the matter density \Omc, and the vacuum energy density \Olc.
(All 68\% confidence limits.)
\end{abstract}


\section{Introduction}

The angular power spectrum of the cosmic microwave
background (CMB) has much to teach us about the nature of the
Universe in which we live~\markcite{hu97b}(Hu, Sugiyama, \& Silk 1997).
Measurements are improving rapidly~\markcite{debernardis00,hanany00,padin01}({de Bernardis} {et~al.} 2000; {Hanany} {et~al.} 2000; {Padin} {et~al.} 2001),
and for a wide variety of theoretical scenarios the predicted
spectra can be calculated accurately~\markcite{zaldarriaga00}({Zaldarriaga} \& {Seljak} 2000).
Comparison of the data and models allows quantitative constraints
to be placed on the parameters of the Universe in which we find
ourselves, and is the subject of this paper.

The Degree Angular Scale Interferometer (\dasi), along with
its sister instrument the CBI~\markcite{padin01}({Padin} {et~al.} 2001) and the
VSA~\markcite{jones96}({Jones} 1996), is one of several new compact 
interferometers specifically designed for observations of the CMB. 
This paper is the third in a set of three.
Paper I~\markcite{leitch01}(Leitch {et~al.} 2001) gives a detailed description of
the instrument, observations, and data calibration.
Paper~II~\markcite{halverson01}(Halverson {et~al.} 2001) focuses on the
extraction of the angular power spectrum from the
calibrated interferometric data, and provides band-power
estimates of the angular power spectrum of the CMB.
In this paper we combine the low-$l$ measurements made
by the COBE-DMR instrument~\markcite{bennett96}({Bennett} {et~al.} 1996) with the new
measurements from \dasi\ to constrain the parameters
of cosmological models.

The layout of this paper is as follows.
The considerations which drive our selection of the
model and parameter space to probe are detailed in \S\ref{sec:grid}.
In \S\ref{sec:datacomp} we review the method used to compare
band power data to theoretical spectra.
The results of this comparison are described in \S\ref{sec:res},
and in \S\ref{sec:conc} we draw some conclusions.

\section{Models, Parameters and Model Grid Considerations}
\label{sec:grid}

Following the discovery of the CMB, and the realization
that the Universe went through a hot plasma epoch,
it was proposed that adiabatic density perturbations
in that plasma would lead to acoustic oscillations~\markcite{peebles70}({Peebles} \& {Yu} 1970), and a series of harmonic peaks in the
angular power spectrum~\markcite{doroshkevich78}({Doroshkevich}, {Zeldovich}, \&  {Sunyaev} 1978).
It was assumed that the initial fluctuations were scale invariant
only because this is the simplest possible case.
It was not until later that (New) Inflation was
proposed~\markcite{guth82,bardeen83,hawking82,starobinskii82}({Guth} \& {Pi} 1982; {Bardeen}, {Steinhardt}, \&  {Turner} 1983; {Hawking} 1982; Starobinsky 1982) ---
an elegant cosmo-genic mechanism which naturally produces such conditions.
The simplest versions of this theory also make the firm prediction
that the Universe is exactly flat, i.e., has zero net
spatial curvature.

Although Inflation sets the stage at the beginning of the
plasma epoch it has nothing to say about the contents
of the Universe.
Over the last several decades a wealth of evidence has
accumulated for the existence of some form of gravitating
matter which does not interact with ordinary baryonic
material; the so-called cold dark matter (CDM).
Conflicting theoretical expectations and experimental measurements
led to the proposal that a third component is present ---
an intrinsic vacuum energy.
This three component model is the scenario we have chosen to consider.

The density required to produce zero net spatial curvature is referred
to as the critical density $\rho_c=3H_0^2/(8\pi G)$, where $H_0$ is the
Hubble constant ($H_0 \equiv 100 h$~km~s$^{-1}$~Mpc$^{-1}$).
It is convenient to measure the present day density of a component
of the Universe in units of $\rho_c$, denoting this $\Omega_i$:
the density of baryonic matter \Ob, the density of CDM \Oc,
and the equivalent density in vacuum energy \Ol.
Thus the density of matter is given by $\Om \equiv \Ob+\Oc$ and the
total density is given by $\Ot \equiv \Om+\Ol$.
Since $\Omega_i=\rho_i/\rho_c$ note that $\Omega_ih^2$ is a
physical density independent of the value of the Hubble constant.

To generate theoretical CMB anisotropy power spectra we have
used version~3.2 of the freely available \cmbfast\
program~\markcite{zaldarriaga00}({Zaldarriaga} \& {Seljak} 2000).
This is the most widely used code of its type, and versions~3
and greater can deal with open, flat and closed universes.
\cmbfast\ calculates how the initial power spectrum
of density perturbations is modulated through the acoustic oscillations
during the plasma phase, by the effects of recombination,
and by reionization as the CMB photons stream through the
Universe to the present.
The program sets up transfer functions, taking as input \Ob, \Oc\ and \Ol,
as well as $H_0$, the optical depth due to reionization (\tc),
and some other parameters.
It can then translate initial perturbation spectra
into the mean angular power spectra of the CMB anisotropy which
would be observed in such a universe today.
Inflation predicts that the initial spectrum is a
simple power law with slope close to, but not
exactly, unity.

In any given comparison of data to a multidimensional
model we may have external information about the
values of some or all of the parameters.
This may come from theoretical prejudice, or from other
experimental results.
It may also be that the data set in hand is unable to simultaneously
constrain all of the possible model parameters to
the precision which we desire.
In such cases we can choose to invoke our external knowledge,
and fold additional information about the preferred parameter
values into the likelihood distribution.
Often this occurs because a parameter which could potentially
be free is fixed at a specific value (an implicit prior).
Or we may multiply the likelihood distribution
by some function which expresses the values of the parameters
which we prefer (an explicit prior).
The choice of measure, e.g., whether a variable is taken
to be linear or logarithmic, is also an implicit prior ---
although such distinctions become less
important as a variable becomes increasingly well constrained.

There is no {\it a priori} reason to suppose that the marginal
likelihoods of the cosmological parameters are Gaussian.
Thus the curvature of the likelihood surface at the peak
does not fully characterize the distribution.
To set accurate constraints it is therefore
necessary to explore the complete likelihood space by
testing the data against a large grid of models.
If the introduction of priors is to be avoided
the grid must be expanded until one can
be confident that it encompasses essentially all of the total likelihood.

Unfortunately it turns out that even within the paradigm of
adiabatic CDM models the anisotropy power spectrum of the CMB does
not fully constrain the parameters of the Universe.
For example, it is well known that $\Om$\ and $\Ol$\ are highly degenerate.
It is always necessary to invoke external information in any
constraint-setting analysis.
The clear articulation of these priors, both implicit and explicit,
is critically important, as has previously been
noted~\markcite{jaffe01}({Jaffe} {et~al.} 2001).

We have chosen to consider a seven dimensional model space.
The parameters which we include are the physical densities of
baryonic matter ($\Obh \equiv \ob$) and CDM ($\Och \equiv \oc$),
as well as the spectral slope of the initial scalar fluctuations
(\ns), and the overall normalization of the power spectrum
as measured by the amplitude at the tenth multipole
($\Cten \equiv l(l+1)C_{10}/2\pi$).
Noting that the degeneracy in the $(\Om,\Ol)$ plane is
along a line of constant total density we opt
to rotate the basis vectors by 45$^\circ$
and grid over the sum and difference of these parameters:
$\Om+\Ol \equiv \Ot$ and $\Om-\Ol \equiv \Od$.
This reduces the size of the model grid required to box
in the region of significant likelihood.
The seventh parameter is the optical depth due to reionization (\tc).
Note that for each point in $(\Od,\Ot,\ob,\oc)$ space
there is an implied value of the Hubble constant
($h = \sqrt{2(\ob + \oc)/(\Ot+\Od)}$)
so we can calculate the $(\Ob,\Oc,\Ol,h)$ values for input to
\cmbfast.

We assume, as is the theoretical prejudice, that the
contribution of tensor mode perturbations is very small 
as compared to scalar, and ignore their effect~\markcite{lyth97}({Lyth} 1997).
Tensor modes primarily contribute power at low $l$ numbers,
so if a large fraction of the power seen by \dmr\ were caused by
this effect the scalar spectrum would need to be strongly
tilted up to provide the observed power at smaller angular scales.
However we know that \ns\ cannot be $\gg1$ as this would
conflict with results from large scale structure studies.
Our constraints should, however, be taken with an
understanding of our assumption regarding tensor modes.

In principle some of the dark matter could be
in the form of relativistic neutrinos (hot as opposed to cold
dark matter).
However the change that this would make to the CMB power spectrum
is negligible compared to the uncertainties of the \dasi\
data~\markcite{dodelson96}({Dodelson}, {Gates}, \&  {Stebbins} 1996).
We therefore assume that all the dark matter is cold,
and set $\Omega_\nu=0$, although it should then be
understood that the \Och\ value we find may, in principle,
contain some hot dark matter.

Papers fitting the BOOMERANG-98 and MAXIMA-1
band-power data~\markcite{balbi00,lange01,jaffe01}({Balbi} {et~al.} 2000; {Lange} {et~al.} 2001; {Jaffe} {et~al.} 2001) considered
seven dimensional grids rather similar to our own.
Other studies have examined model grids with as many as 11
dimensions~\markcite{tegmark01}({Tegmark}, {Zaldarriaga}, \&  {Hamilton} 2001), including an explicit density in neutrinos
and tensor mode perturbations, generally finding both effects to be small.

Taking the philosophy that simplicity is a virtue, we have
not made use of $l$-space or $k$-space splitting
to accelerate the calculation of the model grid~\markcite{tegmark01}({Tegmark} {et~al.} 2001).
In addition we have generated a simple regular grid, rather
than attempting to concentrate the coverage in the
maximum likelihood region.
Finally we have not treated the normalization parameter
\Cten\ as continuous, preferring to explicitly grid over this
parameter as well.
This is computationally somewhat slower, but makes the
marginalization simpler, and involves no assumption about
the form of the variation of $\chi^2$ versus this parameter.
Using the notation lower-edge:step-value:upper-edge (number-of-values)
our grid is as follows: $\Od=-1.0:0.2:3.4$ (23), $\Ot=0.7:0.05:1.3$ (13),
$\Obh=0.0100:0.0025:0.0400$ (13), $\Och=0.00:0.05:0.5$ (11),
$\tc=0.0:0.1:0.4$ (5), $\ns=0.75:0.05:1.25$ (11), and $\Cten=300:50:1300$ (21).
Excluding the small, physically unreasonable corner of parameter
space where $\Om\leq0$, we make $205,205$ runs of \cmbfast\ generating
$205,205 \times 11 = 2,257,255$ theoretical
spectra and calculating $2,257,255 \times 21 = 47,402,355$ values
of $\chi^2$ against the data.

For theoretical and phenomenological discussions of how the various
peak amplitudes and spacings of the power spectrum are related to the
model parameters see \markcite{hu97b}Hu {et~al.} (1997) and \markcite{hu01}{Hu} {et~al.} (2001).
In this paper we choose to compare data and models without
explicitly considering such connections.

\section{Comparison of Data and Models}
\label{sec:datacomp}

Consider a set of observed band-powers $\mathcal{D}_i$ 
in units of $\mu$K$^2$, together with their covariance matrix
$P_{ij}$.
If the overall fractional calibration uncertainty of the experiment is $s$
we can add this to the covariance matrix as follows:
\begin{equation}
N_{ij} = P_{ij} + s^2 \mathcal{D}_i \mathcal{D}_j.
\label{eqn:varcal}
\end{equation}
For the purposes of the present analysis we assume $s=0.08$
which includes both temperature scale and beam
uncertainties (the fractional error on the band-powers is 8\%
corresponding to 4\% in temperature units --- see Paper II).

\begin{figure*}
\plotone{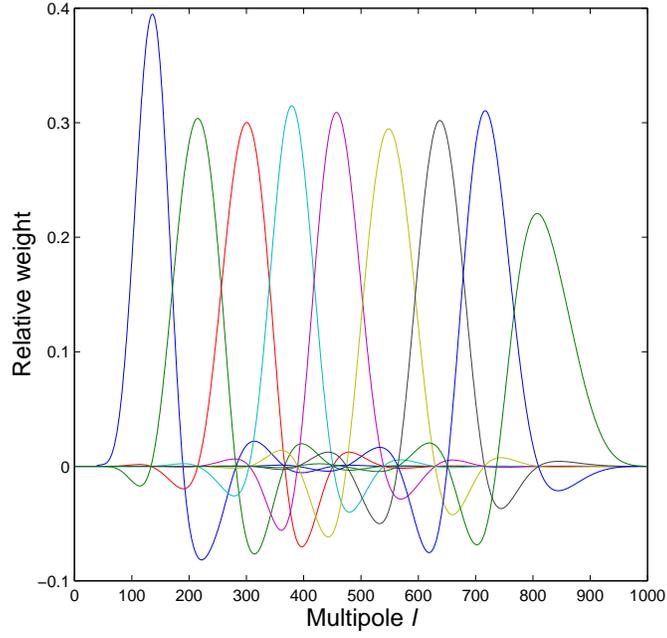}
\caption{Window functions for the \dasi\ band-powers.
\label{fig:wfunc}}
\end{figure*}

Now consider a model power spectrum $\mathcal{C}_l$.
The expectation value of the data given the model is obtained
through the ``band-power'' window function
$W^\mathrm{B}_{il}/l$~\markcite{knox99}({Knox} 1999),
\begin{equation}
\mathcal{T}_i = \sum_l \mathcal{C}_l \frac{W^\mathrm{B}_{il}}{l}.
\label{eqn:wfunc}
\end{equation}
The band-power window functions $W^\mathrm{B}_{il}/l$ are calculated from the
band-power Fisher matrix, $F$, and the Fisher matrix
$F^\mathrm{s}$ of the bands, $b_i$, sub-divided into individual multipole
moments,
\begin{equation}
\frac{W^\mathrm{B}_{il}}{l} = \sum_{i'} {\left(F^{-1}\right)}_{ii'}
	\sum_{l' \in b_{i'}} F^\mathrm{s}_{ll'}, 
\label{eqn:wfcalc}
\end{equation}
\markcite{knox99}(adapted from  {Knox} 1999,  for Fisher matrices with significant
off-diagonal elements).  The sum of each row of the
array $W^\mathrm{B}_{il}/l$ is unity, so Equation~(\ref{eqn:wfunc})
simply represents a set of weighted means.  Note that any experiment
with less than full sky coverage will always have non top-hat 
band-power window functions.
In practice, we calculate Equation~(\ref{eqn:wfcalc}) by
subdividing each band into four sub-bands, and interpolate the results. 
The functions for the \dasi\ band-powers
are plotted in Figure~\ref{fig:wfunc}, and are available at
our website\footnote{\tt http://astro.uchicago.edu/dasi}.  In practice
the effect of using the correct window function, versus simply
choosing the $\mathcal{C}_l$ at the nominal band center, is extremely
modest.

The uncertainties of the $\mathcal{D}_i$ are non-Gaussian so it
would not be correct to calculate $\chi^2$ at this point.
However it is possible to make a transformation such that the uncertainties
become Gaussian to a very good approximation~\markcite{bond00}({Bond}, {Jaffe}, \& {Knox} 2000).
An additional set of quantities $x_i$ need to be calculated from the
data which represent the component of the total uncertainty which
is due to instrument noise.
We can then transform each of the variables as follows,
\begin{eqnarray}
\mathcal{D}_i^Z & = & \ln(\mathcal{D}_i + x_i) \\
\mathcal{T}_i^Z & = & \ln(\mathcal{T}_i + x_i) \\
M_{ij}^Z & = & N_{ij}^{-1}(\mathcal{D}_i + x_i)(\mathcal{D}_j + x_j),
\end{eqnarray}
and calculate $\chi^2$ as usual,
\begin{equation}
\chi^2 = (\mathcal{D}_i^Z - \mathcal{T}_i^Z) M_{ij}^Z
(\mathcal{D}_j^Z - \mathcal{T}_j^Z).
\end{equation}

The inverse covariance matrix elements $M^Z_{ij}$ will be
approximately independent of the bandpowers $\mathcal{D}_i$. This is
true even with the added calibration uncertainty term in
Equation~(\ref{eqn:varcal}), under the assumption that the band-power
uncertainty is sample variance dominated, i.e., $\mathcal{D}_i/x_i
\gg 1$ (as is the case with almost all the \dasi\ band-powers), or that
the fractional calibration uncertainty is small compared to the total
uncertainty in the band-power, $s^2 \ll N_{ii}/\mathcal{D}_i^2$, which
is the case for \dmr. Use of this transformation is very important as
it allows us to use $\chi^2$, and therefore not only to find the best
fitting model, but to determine an absolute goodness of fit.

The ability of smaller angular
scale ($l>100$) CMB data to set constraints on model
parameters is much improved when the large angular scale
($l\leq25$) information from the \dmr\ instrument is included.
We use the \dasi\ bandpowers described and tabulated in Paper~II
together with the 24~\dmr\ band-powers provided in the
RADPACK distribution~\markcite{knox_radpack,bond00}(Knox 2000; {Bond} {et~al.} 2000),
concatenating the $\mathcal{D}_i$ and $x_i$ vectors and
forming a block diagonal covariance matrix.
Note that while the effect of the transformation described
above is modest for the \dasi\
points, it is very important for those from \dmr\
(due to the large sample variance at the lower $l$'s).

\section{Results}
\label{sec:res}

\begin{figure*}
\plotone{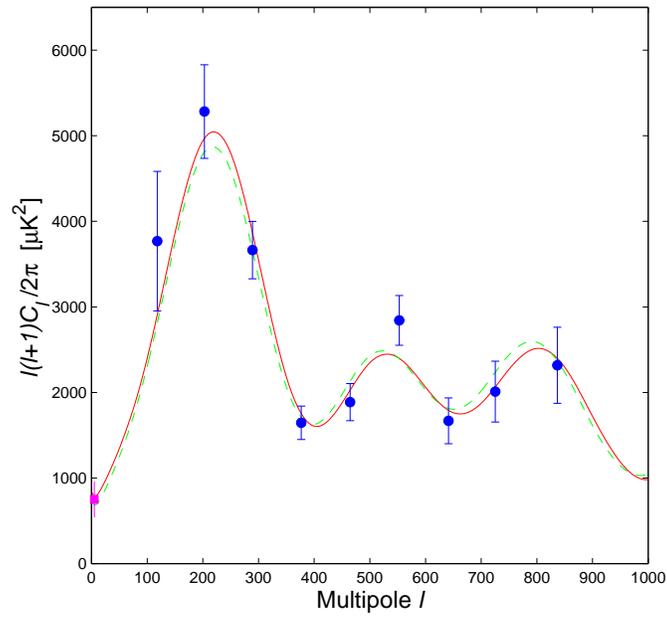}
\caption{The DASI first-season angular power spectrum in
nine bands (closed circles).
The \dmr\ information is shown compressed to the single lowest $l$ point.
The solid (red) line is the best fitting model which falls on our
grid, while the dashed (green) shows the concordance model
$(\Ob,\Oc,\Ol,\tc,\ns,\h) = (0.05,0.35,0.60,0,1.00,0.65)$.
The error bars plotted here are strictly for illustrative
purposes only.
The $\chi^2$ calculation is made using the full covariance matrix,
and after the transformation described in \S\ref{sec:datacomp}.
Thus ``Chi-by-eye'' can be misleading.
\label{fig:bestfit}}
\end{figure*}

Figure~\ref{fig:bestfit} shows the \dasi\ band-powers, together
with the \dmr\ data condensed to a single point for display.
The $\chi^2$ of the best fit model which falls on our grid
is $29.5$ for the 9 \dasi\ plus 24 \dmr\ band-powers.
Assuming a full 7 degrees of freedom are lost to the fit,
this is at the 71\% point of the cumulative distribution
function (cdf)\footnote
{In fact since there is not sufficient freedom in the model
that an arbitrary set of 7 band-powers could be fit exactly
the effective loss of degrees of freedom is less than 7.
For example assuming 4 shifts us to the 56\% point of the
cdf.}.
The parameters of this model are \bfm, equivalent to \bfma.
However, no particular importance should be ascribed to these ---
the concordance model~\markcite{ostriker95,krauss95}({Ostriker} \& {Steinhardt} 1995; {Krauss} \& {Turner} 1995) which is
shown has a $\chi^2$ of 30.8 (76\%) and is also rather a good fit.
The message of Figure~\ref{fig:bestfit} is simply that there
are models within the grid which fit acceptably well,
and that we are therefore justified in proceeding to marginalized
parameter constraints.
We convert $\chi^2$ to likelihood, $\mathcal{L}=e^{-\chi^2/2}$.

The extreme degeneracy of CMB data in the $(\Om,\Ol)$ plane
has already been mentioned.
This inability to choose between models with the same \Ot\ is
in fact weakly broken at very low-$l$ numbers by the Sachs-Wolfe
effect~\markcite{efstathiou99}({Efstathiou} \& {Bond} 1999).
The likelihood contours diverge from the $\Ot=1$ line as
\Od\ becomes $\gg1$, and the allowed region broadens.
Consequently the marginal likelihood curve of
\Ot\ acquires a high side tail as models with progressively
greater \Od\ are included.
These high \Od\ models have very low values of \h, and
are known to be invalid from a wide range of non-CMB data.
This being the case it is clearly not sensible to allow
them to influence our results.

We are therefore prompted to introduce additional external information.
We could simply restrict \Om\ and \Ol\ to some ``reasonable'' range;
for example requiring $\Ol>0$ and $\Om<1$.
Instead we choose to introduce a prior on \h\ since this strongly
breaks the degeneracy and is a quantity which is believed to have
been measured to 10\% precision~\markcite{freedman00}(Freedman {et~al.} 2000).
We use two $h$ priors; a weak $h>0.45$, and a strong $h=0.72\pm0.08$
assuming a Gaussian distribution.
For the weak prior adding the additional restriction $h<0.90$ has
almost no effect as the excluded models already have very small likelihoods.
To apply the prior we simply calculate the \h\ value at each grid
point, assign it the relevant likelihood,
and multiply the two grids together.

\begin{figure*}
\resizebox{\textwidth}{!}{\includegraphics{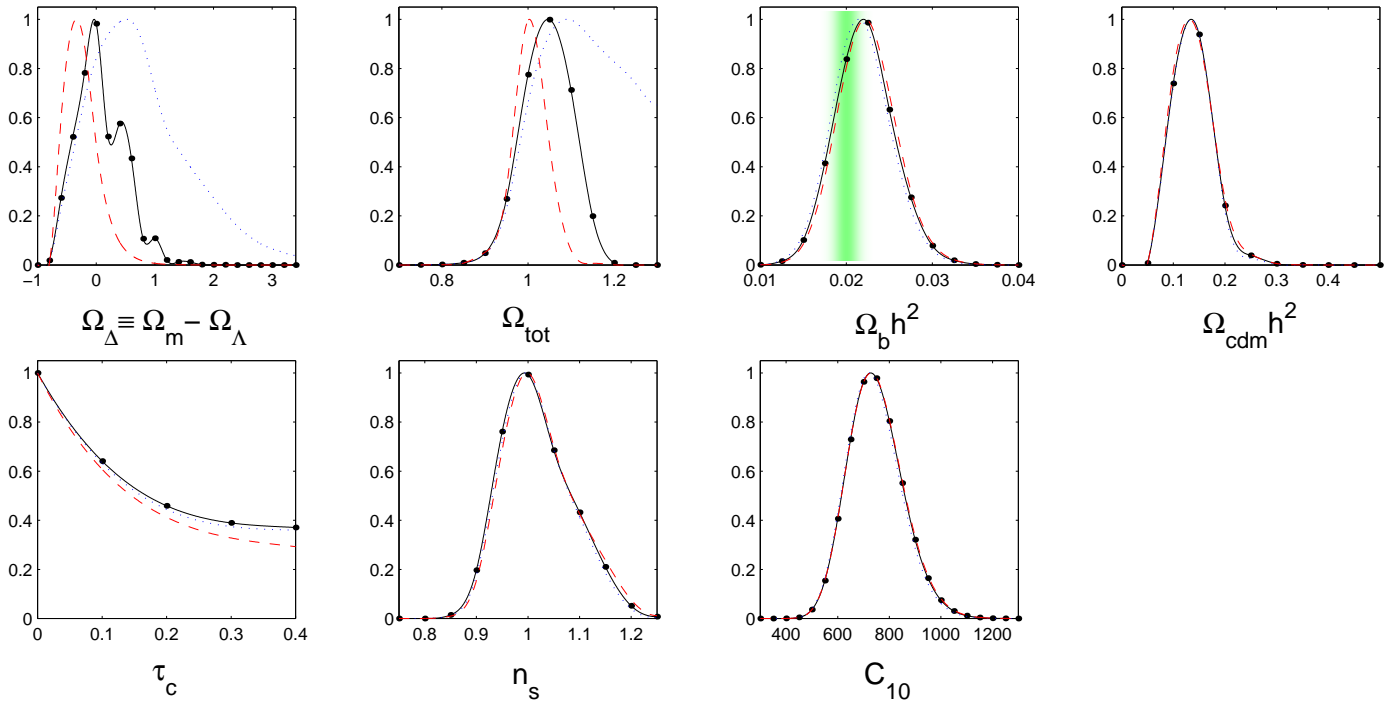}}
\caption{Marginal likelihood distributions for each dimension of
the model grid.
The dotted (blue) and solid (black) lines show the distribution before
and after the introduction of a weak prior on the Hubble parameter
($h>0.45$).
The dashed (red) lines shows the effect of the stronger prior $h=0.72\pm0.08$.
In the \Obh\ panel the BBN constraint is shown as a (green) shaded region.
All curves are normalized to a peak height of unity, and are spline
interpolations of the actual model grid values shown by the points.
\label{fig:marcons1}}
\end{figure*}

Figure~\ref{fig:marcons1} shows how the marginal likelihood distributions
of the model parameters change as we move from the implicit prior
of $(\Od\leq3.4,\Ot\leq1.3)$, to weak, and then strong prior cases.
Note that most of the curves fall to a small fraction of their peak
value before the edge of the grid is reached.
For {\it all} parameters where this is not the case one must
introduce external priors such that it becomes so, and/or
acknowledge the implicit top-hat prior which the range of that
grid parameter represents --- only then can the constraint on
{\it any} of the parameters be accepted.
All such priors must then be quoted with the constraints.
In fact we see that $(\Obh,\Och,\ns,\Cten)$ are almost completely
unaffected by the choice of prior on \h\ --- this indicates that
correlations between these parameters and $(\Od,\Ot)$ are
modest, and is a fortunate result.

We turn now to \tc\ which, as is evident in Figure~\ref{fig:marcons1},
is a very poorly constrained parameter.
The effect of reionization is to suppress power at small angular
scales, and tilt the spectrum down.
However this can be compensated by adjusting \ns\ upwards making
these two parameters largely degenerate.
Worse still, unlike the \h\ prior discussed above, we have
very weak experimental guidance as to the value
of \tc\ --- we know only that reionization occurred at $z\geq6$,
roughly equivalent to $\tc\geq0.03$.
From theoretical ideas regarding early structure
formation, and energy emission, it seems essentially impossible
that $\tc>0.4$ \markcite{haiman99}(see Haiman \& Knox 1999,  for a recent review of our
knowledge regarding reionization).
The theoretical prejudice for $\ns\approx1$ is strong, but since
this is a fundamental parameter of the cosmology which we are
trying to measure, we are very reluctant to place a prior
on it.

We have opted to accept the top-hat prior implicit in our
model grid, i.e., that the likelihood of \tc\ falling between
0.0 and 0.4 is uniform.
In Table~\ref{tab:marcons1} we list the spline interpolated
median, $1\sigma$ and $2\sigma$ points of the integral
constraint curves.
The modal (maximum likelihood) value is also given.
All of the constraints quoted in this paper are 68\% confidence
intervals about the median.
Although one can argue that the mode is perhaps more natural,
in practice it makes little difference.

\begin{table*}
\caption{Parameter constraints from \dasi+\dmr\ data
\label{tab:marcons1}}
\begin{center}
\begin{tabular}{|l|l|l|l|l|l|l|}
\tableline
Parameter & 2.5\% & 16\% & 50\% & 84\% & 97.5\% & mode \\
\tableline
\Ot   &  0.927 &  0.986 &  1.044 &  1.103 &  1.150 &  1.047 \\ 
\Obh  & 0.0156 & 0.0187 & 0.0220 & 0.0255 & 0.0292 & 0.0220 \\ 
\Och  &  0.075 &  0.100 &  0.137 &  0.175 &  0.225 &  0.135 \\ 
\ns   &  0.901 &  0.949 &  1.010 &  1.092 &  1.166 &  0.993 \\ 
\Cten &    558 &    642 &    741 &    852 &    973 &    728 \\
\tableline
\end{tabular}
\end{center}
These constraints are from a seven dimensional
grid, assuming the weak prior $h>0.45$, and \tcp.
The indicated points on the cumulative distribution function are given,
as well as the maximum likelihood value.
\end{table*}

Referring to Figure~\ref{fig:marcons1} we see that the \Od\
constraint is almost wholly driven by the prior on \h;
for this reason we have not included it in table~\ref{tab:marcons1}.
However, if one is prepared to accept the strong prior
$h=0.72\pm0.08$, then our data indicates that
\Otcsh, \Omc\ and \Olc.

\begin{figure*}
\resizebox{\textwidth}{!}{\includegraphics{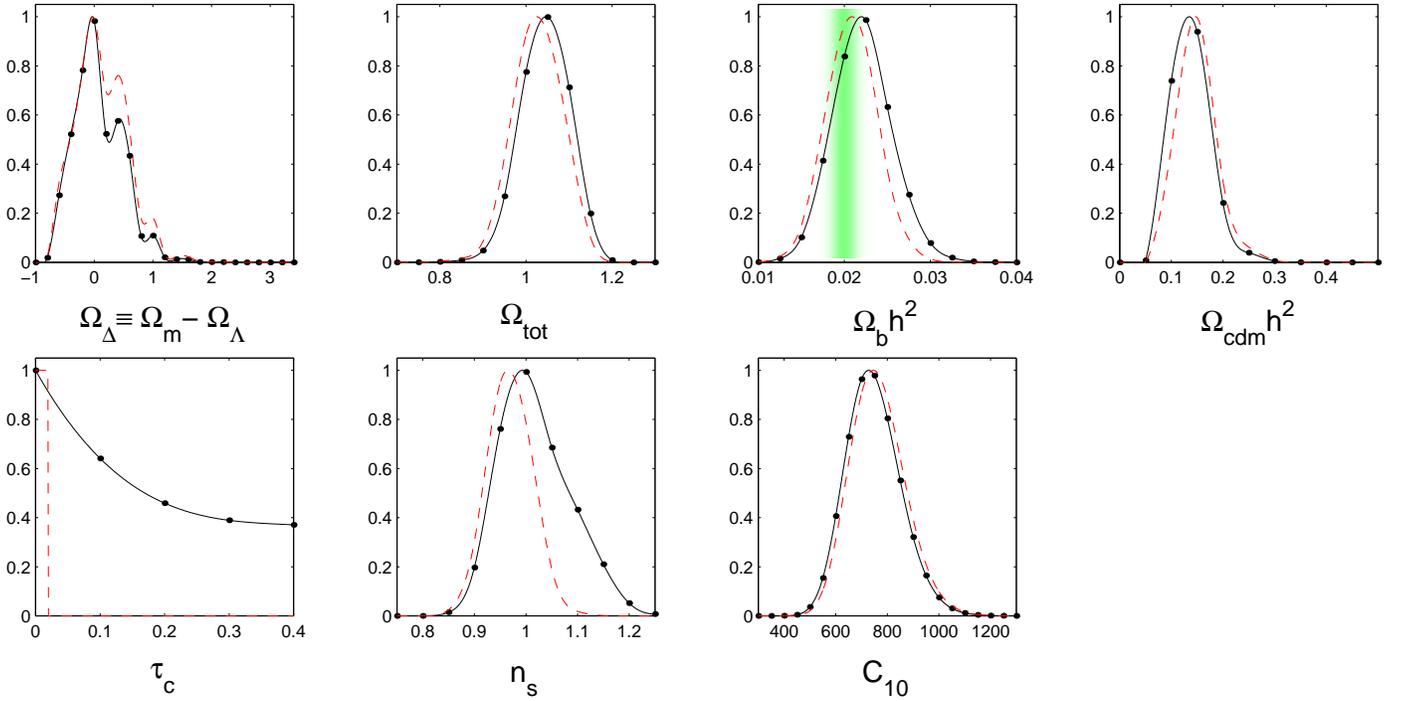}}
\caption{Marginal likelihood distributions when varying the
prior on \tc.
All curves assume the weak prior $h>0.45$.
The solid (black) are the same as in Figure~\ref{fig:marcons1} and
assume \tcp, while the dashed (red) set $\tc=0.0$.
\label{fig:marcons2}}
\end{figure*}

Figure~\ref{fig:marcons2} illustrates the effect of setting $\tc=0.0$
(which the data weakly prefers).
As already mentioned the primary degeneracy is with \ns\
which shifts to \nsct.

The selection of the particular set of nine bandpowers which we have
used in this analysis is quite arbitrary.
We have tested increasing the number of bins --- as expected the
variance and covariance of the points increases to compensate, and
the constraint curves remain the same.
Shifting the bins in $l$ also leaves the results unchanged ---
for instance the alternate binning shown in Paper~II
leads to results which are indistinguishable from those
presented here.

\section{Conclusion}
\label{sec:conc}

We have compared the \dasi+\dmr\ data to adiabatic CDM models
with initial power law perturbation spectra.
The best fitting model has an acceptable $\chi^2$ value,
indicating that for data of the present quality models within this
class are a plausible representation of the underlying physics.
Adopting the conservative priors $h>0.45$ and \tcp,
we find \Otc\ and \nsc, consistent with the predictions
of Inflation.
Moving to more aggressive priors on \h\ and \tc\ tightens the
constraints on \Ot\ and \ns\ respectively but they remain
consistent with the theory.

We find that \Obhc\ and \Ochc, adding to the already
very strong evidence for non-baryonic dark matter.
These constraints are only weakly affected by
the choice of \h\ and \tc\ priors.
Setting a strong \h\ prior breaks the $(\Om,\Ol)$ degeneracy
such that we constrain \Omc\ and \Olc --- consistent with other
recent results.

The current best value for \Obh\ derived from the well developed
theory of big bang nucleosynthesis (BBN), 
combined with measurements of the primeval deuterium abundance, is
$\Obh=0.020\pm0.002$~\markcite{burles01}(Burles, Nollett, \& Turner 2001,  95\% confidence).
The $\chi^2$ of the difference between this and our
own value is at the 42\% point of the cdf (assuming Gaussian errors
on both); the values are hence consistent.
Previous CMB analyses have seen little power in the
second peak region, and have determined \Obh\ values
higher than, and inconsistent with, BBN at the $\approx3\sigma$
level~\markcite{lange01,jaffe01}({Lange} {et~al.} 2001; {Jaffe} {et~al.} 2001).

\acknowledgments

We would like to thank the staff of Argonne National Laboratory for allowing
us to use the ``Chiba City'' computer cluster to generate
the model grid used in this paper.
Lloyd Knox, Wayne Hu and Mike Turner are thanked for useful conversations.
Most importantly we thank U.\ Seljak \& M.\ Zaldarriaga for making
\cmbfast\ publicly available.
We would also like to thank the CBI team at Caltech
for assistance with the design and implementation of the \dasi\ instrument.
This research is supported by the National Science Foundation under a
cooperative agreement (NSF OPP 89-20223) with CARA, a
National Science Foundation Science and Technology Center.

\bibliography{}

\end{document}